\documentstyle[11pt,aaspp4]{article}

\begin{document}

\title{Could a Local Group X-Ray Halo Affect the X-Ray and Microwave
Backgrounds?}

\author{Rachel A. Pildis}
\affil{Harvard-Smithsonian Center for Astrophysics, 60 Garden St., MS 83,
Cambridge, Massachusetts 02140; \\rpildis@cfa.harvard.edu}

\and

\author{Stacy S. McGaugh}
\affil{Department of Terrestrial Magnetism, Carnegie Institution of
Washington, 5241 Broad Branch Road, NW, Washington, DC 20015;
\\ssm@dtm.ciw.edu}

\begin{abstract}
Suto et al.\ (1996, \apjl, 461, L33) have suggested that an X-ray halo in
the Local Group might explain both the observed low-energy excess in the
X-ray background and the quadrupole anisotropy in the cosmic microwave
background.  Recent observations of poor groups of galaxies by the $ROSAT$
PSPC set reasonable limits on how extensive and dense such a halo could be.
The poor groups most similar to the Local Group do not have a detectable
halo, and the upper limits of these observations suggest that any Local
Group halo would be nearly two orders of magnitude too tenuous to produce
the effects Suto et al.\ (1996) discuss.  In particular, the
Sunyaev-Zel'dovich effect cannot contribute significantly to the quadrupole
anisotropy measured by COBE.
\end{abstract}


\newpage

\section{Introduction}

	Hot, X-ray--emitting gas is an important component of the gaseous
content of the universe.  It usually dominates the interstellar medium of
elliptical galaxies (\cite{fab89,bre92}).  It also accounts for most of the
baryonic mass of rich clusters of galaxies and up to 30\% of their total
gravitating mass (\cite{whi93,elb95}).  Such observations led Suto et
al.\ (1996; hereafter SMIO) to suggest that the Local Group of galaxies
might have a substantial X-ray halo of its own.

	This hypothesized hot intragroup medium (IGM) could contribute much
of the excess in the soft ($\sim$1 keV) X-ray background.  Recent
measurements have shown that this excess is well-fit by a thermal model with
solar metallicity and a temperature of 0.16 keV (\cite{gen95}).  Determining
the contributors to the X-ray background is crucial to understanding it.
Furthermore, SMIO point out that such a Local Group X-ray halo would produce
a significant quadrupole anisotropy in the microwave background through the
Sunyaev-Zel'dovich (SZ) effect (\cite{zel69}).  This quadrupole signature,
$Q_{SZ}$, could in principle account for the entirety of the quadrupole
signal observed by COBE:  $Q_{COBE} \approx 6 \mu$K (e.g.,
\cite{gor94,ben94}).

	The COBE measurement provides a critical constraint on cosmological
models.  The power spectrum of primordial density fluctuations is usually
parameterized as $P(k) = A k^n$.  The spectral index $n$ is a primary
component of any cosmological model, with the scale invariant value $n=1$
being the preferred ``natural'' value of inflationary models.  The COBE
observations provide both the normalization $A$ and powerful constraints on
$n$.  The COBE normalization on large angular scales is far too high
relative to the power on smaller (galaxy to cluster) scales for the
predictions of the ``standard'' $n=1$ cold dark matter model (e.g.,
\cite{fis93}).  The need for relatively more power on large scales as
indicated by COBE is a primary motivation for considering contrived models
with admixtures of hot as well as cold dark matter.  A significant local
contribution to $Q_{COBE}$ through the SZ effect would alter the range of
viable cosmological models significantly, perhaps even reviving standard
cold dark matter.  It is therefore very important to place limits on the
potential $Q_{SZ}$ of the putative Local Group hot IGM.

      Since we are embedded in the hot interstellar medium of the Galaxy,
it is exceedingly difficult to disentangle the possible contribution of a
hot IGM to the diffuse X-ray background from other, more local sources.
However, the Local Group can be placed in context of observations of other
groups of galaxies.  While the suggestion of a significant Local Group IGM
was made as a parallel to the highly luminous intracluster media of rich
clusters, groups provide a more enlightening comparison.  With the advent of
the $ROSAT$ PSPC, many observations of galaxy groups of varying richness
have been obtained, so such comparisons can be made.

\section{The IGM in Other Groups}

	While poor groups of galaxies tend to be somewhat more populated and
more dense (in projection) than the Local Group, they provide a much better
comparison for the X-ray properties of the Local Group than rich clusters.
The $ROSAT$ PSPC has provided the necessary sensitivity and angular
resolution to make a detailed study of the diffuse X-ray emission seen in
some galaxy groups.  PSPC observations reveal that, in general,
X-ray--luminous groups contain more elliptical galaxies than spirals and the
first-ranked galaxy in these groups is an elliptical (Pildis, Bregman, \&
Evrard 1995, \cite{mul96}).   Very few groups with a spiral majority (the
Local Group has no bright ellipticals) have detected X-ray emission, perhaps
due to their not being sufficiently compact to contain an IGM or the
temperature of the IGM being too low for the PSPC to detect.

\begin{deluxetable}{lcccccc}
\tablenum{1}
\tablewidth{0pt}
\tablecaption{Density and temperature parameters for model and observed groups}
\tablehead{ \colhead{Group} & \colhead{Type}
& \colhead{$r_c$ (kpc)} & \colhead{$n_0$ (cm$^{-3}$)}
& \colhead{$k$T (keV)} & \colhead{$x_0$/$r_c$}
& \colhead{$n_0 r_c$T$_{\rm keV}$ (cm$^{-2}$)} }
\startdata
Suto et al.\ & \nodata & 150 & $1 \times 10^{-4}$ & 1 & 2.3
& $4.6 \times 10^{19}$ \nl
\tablevspace{10pt}
HCG 12 & E & 45 & $5.7 \times 10^{-3}$ & 0.72 & 7.8 & $5.7 \times 10^{20}$ \nl
HCG 62 & E & 60 & $1.5 \times 10^{-3}$ & 1.1 & 5.8 & $3.1 \times 10^{20}$ \nl
HCG 68 & E & 33 & $1.3 \times 10^{-3}$ & 0.98 & 10.6 & $1.3 \times 10^{20}$ \nl
HCG 97 & E & 8 & $6.0 \times 10^{-3}$ & 0.97 & 43.8 & $1.4 \times 10^{20}$ \nl
N2300 group & E/S & 27 & $2.8 \times 10^{-4}$ & 0.93 & 13.0
& $2.2 \times 10^{19}$ \nl
\tablevspace{10pt}
HCG 2  & S & 33 & $<3.8 \times 10^{-4}$ & 1 & 10.6 & $<3.9 \times 10^{19}$ \nl
HCG 10 & S & 33 & $<4.4 \times 10^{-4}$ & 1 & 10.6 & $<4.5 \times 10^{19}$ \nl
HCG 44 & S & 33 & $<2.2 \times 10^{-4}$ & 1 & 10.6 & $<2.2 \times 10^{19}$ \nl
HCG 79 & E & 33 & $<3.5 \times 10^{-4}$ & 1 & 10.6 & $<3.6 \times 10^{19}$ \nl
HCG 93 & S & 33 & $<5.4 \times 10^{-4}$ & 1 & 10.6 & $<5.5 \times 10^{19}$ \nl
\enddata
\tablecomments{Data from Pildis et al.\ (1995) and model from Suto et al.\
(1996).  For the purposes of this comparison, we retain the value of $x_0$
(Milky Way to group center distance) assumed by Suto et al.\ (1996).
Group type indicates whether the group is spiral or
elliptical dominated; the NGC 2300 group contains one bright galaxy of each
type.  Groups with upper limits have their core radius and temperature
fixed, as discussed in the text.}
\end{deluxetable}

        Since over two dozen galaxy groups have been detected with the PSPC,
their properties can be used to constrain models of the possible IGM of the
Local Group.  The relevant quantities are T, the temperature of the plasma,
the column density, which one can parameterize as $n_0 r_c$ (where $n_0$ is
the central electron density and $r_c$ is the core radius), and the distance
of the Milky Way from the center of the Local Group, $x_0$.  SMIO assumed a
Local Group IGM with T=1 keV, $n_0$=$10^{-4}$ cm$^{-3}$, and $r_c$=150 kpc,
and assumed $x_0$=350 kpc.  Table 1 lists the values for these quantities
for the groups analyzed by Pildis et al.\ (1995), as well as the values that
SMIO assumed.  For the purposes of this comparison, we retain the value of
$x_0$ used by SMIO, as if one were to replace the Local Group IGM with that
of the other groups.  Three groups in the sample of Pildis et al.\ (1995)
are not included in Table 1 because they are not spatially resolved in the
PSPC observations:  HCG 4, 92, and 94 (see \cite{pil95} for details).  The
upper limits assume the typical core radius (33 kpc) and temperature (1 keV)
observed in the X-ray--luminous groups, and use a calculated central density
determined from the 3$\sigma$ upper limit on the number of counts within the
group radius in their PSPC observations (\cite{pil95}).  Mulchaey et
al.\ (1996) do not list central densities, but find average temperatures of
$\sim$1 keV and core radii ranging from 9 kpc to over 300 kpc, with a median
value of 43 kpc (excluding the groups from \cite{pil95}).  Note that
the observed groups tend to have smaller values of $r_c$ and greater values of
$n_0$ than SMIO assumed for the Local Group IGM, while
the values for T are similar.

\begin{figure}
\vbox{
\hbox{\includegraphics{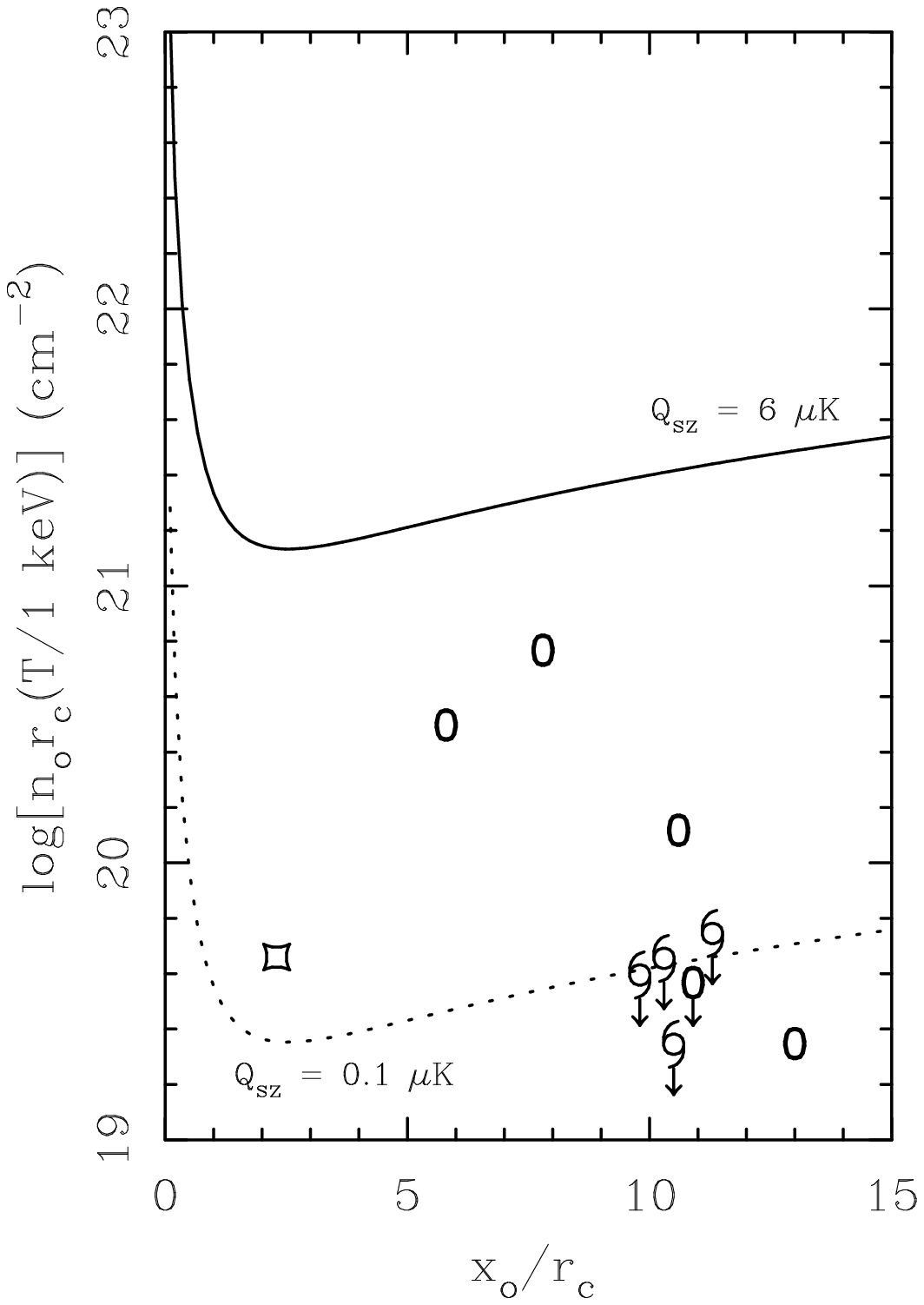}}
\vskip 6.0in
\caption{The density, size, and temperature of the IGM of poor groups of
galaxies analyzed in Pildis et al.~(1995) compared to quadrupole
anisotropies of the microwave background, after Fig.\ 4 of Suto et
al.\ (1996).   The 6 $\mu$K anisotropy curve is taken from Suto et
al.\ (1996) and the 0.1 $\mu$K curve is a scaled version of the 6 $\mu$K
curve.  Ovals mark elliptical-dominated groups and spirals mark
spiral-dominated groups, and downward-pointing arrows indicate upper
limits.  For clarity, the groups with upper limits are scattered around the
assumed value of 10.3 for $x_0$/$r_c$. Also shown, as a square, is the Local
Group X-ray halo model assumed by Suto et al.\ (1996). Note that no
groups---even the model of Suto et al.---come close to the 6 $\mu$K curve,
including those that are richer and more elliptical-dominated than the Local
Group.}
}
\end{figure}

	Figure 1 shows an expanded version of Fig.~4 of SMIO,
with the addition of the data presented in Table 1 and SMIO's own
model for the Local Group IGM.
The theoretical curves follow from equation~(9) of SMIO,
\begin{equation}
n_0 r_c = \frac{4f}{\sqrt{5} \pi} Q_{SZ} \frac{m_e c^2}{\sigma_T kT}
\left(\frac{x_0}{r_c}\right) \left[\tan^{-1}\left(\frac{x_0}{r_c}\right)
-3\left(\frac{x_0}{r_c}\right)^{-1} +3\left(\frac{x_0}{r_c}\right)^{-2}
\tan^{-1}\left(\frac{x_0}{r_c}\right)\right]^{-1},
\end{equation}    
where $m_e$ is the electron mass, $c$ the speed of light, $\sigma_T$ the 
Thomson scattering cross section, and $k$ is the Boltzmann constant. 
There is also a numerical fudge factor $f$ due to the spherical harmonic
multipoles (see equations 4 and 5 of SMIO) which SMIO
calculate numerically but do not explicitly state.  We have adopted 
$f = 8.7$ to match their curve for $Q_{SZ} = 6~\mu$K. 
One detected group, HCG 97, does not appear
on the graph due to its large value of $x_0$/$r_c$.  Note that the
NGC 2300 group, which is dominated by neither spirals nor ellipticals, has
the lowest value of $n_0 r_c$T of all the detections.

	Even the brightest elliptical-dominated groups in the Pildis et
al.\ (1995) sample fall well short of producing the observed $Q_{COBE}
\approx 6\mu$K.  Also well below the curve, puzzlingly enough, are the IGM
parameters assumed by SMIO.  It is rather mysterious why they explicitly
suggest that ``the LG X-ray halo can potentially have a significant effect
on the quadrupole of the CMB anisotropies'' when their own optimistic
estimate of the Local Group parameters give $Q_{SZ} \ll Q_{COBE}$.  The
maximum quadrupole anisotropy consistent with the spiral-dominated groups
appears to be 0.1 $\mu$K, a factor of 60 less than $Q_{COBE}$.  So, unless
the Local Group is extremely unusual, any hot IGM that it may contain is
simply not relevant to the COBE quadrupole anisotropy measurement.

\section{Discussion}

	While it is clear that many systematic effects come into play in
determining the characteristics of an IGM from an X-ray observation (notably
the assumed metallicity of the gas and the background level of the
observation; see \cite{mul96} for an enlightening discussion), these effects
could change the core radius and central density fitted to these groups by
no more than a factor of a few, and likely much less.  The temperature of
galaxy groups are well-determined to be close to 1 keV on average.  With the
further observation that the Local Group is less populated and more diffuse
than even the average {\it undetected} spiral-dominated group, it appears
extremely unlikely that a Local Group IGM could produce a significant
fraction of the COBE quadrupole anisotropy.

	Perhaps a more plausible source for the soft excess in the X-ray
background is a combination of the probable X-ray halo of the Milky Way plus
emission from the Local Superbubble.  Previous studies of the 1/4 and 3/4
keV X-ray backgrounds point towards their origination in the Galaxy rather
than the Local Group (\cite{mcc90}).  Recent observations of X-ray shadowing
by high-latitude molecular clouds demonstrates that the 3/4 keV background
is dominated by emission from the Local Superbubble, while the 1/4 keV
background is likely to arise from a patchy hot halo surrounding the Milky
Way (\cite{sno91,bur91,sno93}).  A significant fraction of spiral galaxies
considered to be similar to our own Galaxy are also seen to have patchy
soft X-ray haloes (\cite{bre94,vog96}).  Since the X-ray emission in
spiral-dominated groups is due solely to the component galaxies, there is no
reason to expect that a Local Group IGM would be an important component of the
local X-ray background.

\section{Conclusion}

	In light of the possibly substantial effect a Local Group
X-ray--emitting halo could have on the local soft X-ray and microwave
backgrounds, as suggested by SMIO, we have considered the
X-ray properties of poor groups of galaxies.  Sparse, spiral-dominated
groups like the Local Group have yet to be detected in X-rays, and even
elliptical-dominated groups have low luminosities relative to rich
clusters.  We therefore conclude that

\begin{enumerate}
\item Unless the Local Group is exceedingly unusual for groups of its type,
it does not have a significant 1 keV X-ray halo.
\item The contribution of any hot gas in the Local Group IGM to the soft (1
keV) X-ray background is negligible.  A more important contributor might be
hot gas associated with the Milky Way itself.
\item The contribution of a Local Group X-ray halo to the COBE quadrupole
anisotropy measurement through the Sunyaev-Zel'dovich effect is completely
negligible:  $Q_{SZ} < 0.1 \mu$K $\ll Q_{COBE} \approx 6 \mu$K.
\end{enumerate}

	Since spiral-dominated groups as yet have only upper limits rather
than actual detections of their IGM, this upper limit on $Q_{SZ}$ is quite
hard unless the Local Group IGM is very exceptional (e.g., comparable to
that of a rich cluster of galaxies).  Thus, the quadrupole anisotropy
measured by COBE must represent the true cosmic signature.  The level at
which this places the normalization of the primordial power spectrum remains
an uncomfortable fact which cosmological models must fit.

\end{document}